**Light-Assisted and Gate-Tunable Oxygen Gas Sensor based on Rhenium Disulfide (ReS$_2$) Field-Effect Transistors**

*Amir Zulkefli, Bablu Mukherjee, Ryoma Hayakawa, Takuya Iwasaki, Shu Nakaharai\*, Yutaka Wakayama\**

A. Zulkefli, Dr. B. Mukherjee, Dr. R. Hayakawa, Dr. T. Iwasaki, Dr. S. Nakaharai
Prof. Y. Wakayama
International Center for Materials Nanoarchitectonics (MANA),
National Institute for Materials Science (NIMS), 1-1 Namiki, Tsukuba 305-0044, Japan.
E-mail: WAKAYAMA.Yutaka@nims.go.jp, NAKAHARAI.Shu@nims.go.jp

A. Zulkefli, Prof. Y. Wakayama
Department of Chemistry and Biochemistry, Graduate School of Engineering, Kyushu University, 744 Moto-oka, Nishi-ku, Fukuoka 819-0395, Japan.

Dr. T. Iwasaki
International Center for Young Scientists (ICYS),
National Institute for Materials Science (NIMS), 1-1 Namiki, Tsukuba 305-0044, Japan.

Gas sensors based on transition metal dichalcogenides (TMDCs) have attracted much attention from a new perspective involving light-assisted or gate-voltage operation. However, their combined roles as regards the gas sensing performance and mechanism have not yet been understood due to the lack of controlled studies. This study systematically investigates the oxygen sensor performance and mechanism of few-layer-thick rhenium disulfide (ReS$_2$) field-effect transistors (FETs) under light illumination and gate biasing. As a result, a combination of light illumination and positive gate voltage enhanced the device responsivity over 100% at a 1% oxygen concentration, that is, the approach achieved a practical sensitivity of 0.01% ppm$^{-1}$, which outperform over most of the reports available in the literature. Furthermore, the fabricated devices exhibited long-term stability and stable operation even under humid conditions, indicating the ability of the sensor device to operate in a real-time application. These results contribute to the development of versatile tunable oxygen sensors based on TMDC FETs.

MAIN TEXT

Oxygen sensors are essential for a wide range of applications related to, for example, medicine, automobiles, food, and the environment.[1–6] Over the past century, sensors have verified oxygen concentrations by using chemical, optical, or electrical approaches.[7,8] Of these approaches, the field-effect transistor (FET) has been widely used in gas sensor applications.[9–11] This is because of its easy incorporation into integrated circuits and well-developed fabrication process. Moreover, current modulation by gate biasing and room-temperature operation provide the advantages of a long device lifetime and low power consumption.[12]

Currently, two-dimensional (2D) materials are being intensively studied for use in gas sensor devices because of their chemical, physical, optical, and electronic features.[13–16] In other words, 2D materials are promising platforms for gas sensing devices thanks to their naturally high surface-area-to-volume ratio, which results in high sensitivity. In particular, transition metal dichalcogenides (TMDCs) have a finite bandgap in the 0.2 to 3.0 eV range depending on the constituent materials and their numbers of layers,[17–19] and they could be used for an FET-based gas sensor.

Although the enhancement of the gas sensitivity of TMDC-based FETs by light illumination and gate biasing[20–25] has been reported, the mechanism and roles of their combined effects are not yet understood. In this regard, among the TMDCs, rhenium disulfide ($ReS_2$) is a promising candidate for a light-assisted and gate-tunable oxygen sensor. This is because $ReS_2$ offers direct bandgap matching with the visible light range regardless of thickness unlike other TMDCs.[26–28] Therefore, $ReS_2$ is an ideal material that offers an optimal trade-off relation between the surface-area-to-volume ratio (thin layer required) and the charge carrier density with the optical absorption (thick layer required) toward realizing a high-performance gas sensor based on a light-assisted and gate-bias operation.

In this study, we investigated $ReS_2$ FET-based oxygen sensor devices. The sensing performance including responsivity, sensitivity, stability, and durability were studied in a systematic manner. We focused particularly on the light illumination and gate voltage dependence to clarify their roles in the sensing mechanism. As a result, a practical sensitivity of 0.01 ppm$^{-1}$ was achieved for oxygen gas by combining light illumination and a positive gate voltage, which outperform over previous reports.[29,30] This research also contributes to an in-depth understanding of the roles of light illumination and gate biasing, leading to the development of a high-performance gas sensor based on TMDC FETs.

**Figure 1**a shows a home-made chamber setup (dimensions: 9.0 cm×6.0 cm×2.5 cm) that we used for the ReS$_2$-FET-based oxygen sensor measurements (see the detailed preparation, fabrication, and sensing measurements of the sensor in Experimental Section). Figure 1b shows a Raman spectrum of the ReS$_2$ nanosheet. Two major peaks were observed at 162.7 cm$^{-1}$ and 214.4 cm$^{-1}$, which represented the E$_g$ mode (in-plane vibrations) and A$_g$-like mode (out-of-plane vibrations), respectively. This result confirmed the high crystallinity of the ReS$_2$ nanosheet.[31] Figure 1c shows an atomic force microscopy (AFM) image and the height profile (green line in the AFM image) of the ReS$_2$ nanosheet. The height profile confirmed that the ReS$_2$ thickness was approximately 6 nm, corresponding to 8-layer-thick ReS$_2$. The 8-layer ReS$_2$ nanosheet was used for all the oxygen sensing measurements in this study because this thickness was found to be optimal (see **Figure S1** for the thickness-dependent properties).

Initially, the transfer curve was measured in a dark condition without oxygen exposure as a reference as shown in **Figure 2**a. A typical n-type operation was observed, i.e., electrons were majority carriers in the ReS$_2$ FET. Then, the influence of light illumination on oxygen sensing was examined. Figure 2b and 2c show the output characteristics of the ReS$_2$ FET in dark conditions and under light illumination, respectively, while no gate voltage was applied. As seen in Figure 2b, the drain current gradually decreased as the oxygen concentration increased from 0 to 10,000 ppm. These results were caused by the nature of the oxygen molecule as an electron acceptor.[32] That is, electrons were transferred from the ReS$_2$ channel surface to the physically adsorbed oxygen molecules to increase the device resistance.

This tendency was enhanced by light illumination as shown in Figure 2c; the variations in the drain current were larger than those obtained under dark conditions. For a clear comparison, the responsivities under both dark conditions and light illumination are plotted as a function of oxygen concentration in Figure 2d, where the oxygen responsivity was determined as the resistance ratio ($\Delta R/R_{N2}$) based on the following equation:

$$\frac{\Delta R}{R_{N2}} = \left[\frac{R_{N2} - R_{O2}}{R_{N2}} \times 100\right]\% \qquad (1)$$

where $R_{N2}$ and $R_{O2}$ are the channel resistances in nitrogen gas (oxygen concentration = 0) and oxygen gas, respectively. The responsivity under dark conditions (black line) showed a slightly monotonical increase from 4% to 19% with increasing oxygen concentration. Meanwhile, light illumination yielded a responsivity of 60% at an oxygen concentration of 10,000 ppm, which is three times higher than that measured under dark conditions. This improvement was achieved by the photo-generated carriers.[33] That is, light illumination increased the electron population in the ReS$_2$ FET, and these electrons were available to interact with oxygen gas molecules. This result clearly demonstrates the advantage of light-assisted gas sensing. For these experiments, the light intensity was fixed at 8.4 mW/cm$^2$ because the maximum responsivity was observed at this intensity (see **Figure S2** for more details).

To examine the effect of the gate voltage on the oxygen sensing performance, the output characteristics were measured under dark conditions with negative ($V_G = -20$ V) and positive ($V_G = +20$ V) gate voltages as shown in **Figure 3**a and 3b, respectively. As can be seen, the variations in the drain current with a negative gate voltage were marginal. Meanwhile, a clear modulation was observed in the drain current with a positive gate voltage. This tendency is further confirmed by the response-oxygen concentration curves shown in Figure 3c. Here, the curve without a gate voltage ($V_G = 0$, black line) is duplicated from Figure 2c for comparison. The responsivity was increased by up to 38% by a positive gate voltage, while a negative gate voltage reduced it to 13% at a 10,000-ppm oxygen concentration. The results can also be explained in terms of the electron population at the ReS$_2$/SiO$_2$ interface. The negative gate voltage depleted the electrons at the ReS$_2$/SiO$_2$ interface. As a result, oxygen gas molecules, which are electron acceptors, receive fewer electrons from the ReS$_2$ FET. In the meantime, the electrons accumulated at the ReS$_2$/SiO$_2$ interface under a positive gate voltage to promote electron transfer with oxygen gas molecules, which contributed to the increment in oxygen gas responsivity.

Next, we examined the combined effect of a positive gate voltage and light illumination on sensing performance. **Figure 4**a shows the output characteristics of the ReS$_2$ FET measured with a combination of light illumination and a positive gate voltage ($V_G$ = +20 V). We observed a distinct decrease in the drain currents with increasing oxygen concentration. Figure 4b summarizes responsivities as a function of oxygen concentration. The ReS$_2$ FET operated under light illumination and a positive gate voltage exhibited improved oxygen gas responsivity, which exceeded 100% at an oxygen concentration of 10,000 ppm. Namely, we achieved a sensitivity of 0.01% ppm$^{-1}$. This is higher than the value obtained in the dark with (0.003% ppm$^{-1}$) and without (0.002% ppm$^{-1}$) a gate voltage, and solely with light illumination (0.005% ppm$^{-1}$) (see **Figure S3** for more details)..

As mentioned above, the gas responsivities are strongly related to the electron population in the ReS$_2$ FET. These phenomena can be further understood from the band diagrams shown in **Figure 5**. These band diagrams illustrate the interaction of the electrons in the conduction band of ReS$_2$ with oxygen gas on the ReS$_2$ surface. Under dark conditions without gate biasing (Figure 5a), the responsivity is predominantly influenced by the doping effect of the gas molecules.[34] Meanwhile, the positive gate biasing leads to electron accumulation in ReS$_2$, which decreases $R_{N2}$ (Figure 5b). Then, the ReS$_2$ surface attracts the oxygen molecules, resulting in an increase in $\Delta R$. Thus, the responsivity ($\Delta R/R_{N2}$) is enhanced. On the other hand, the light illumination also generates electrons by photo excitation, leading to the enhancement of $\Delta R/R_{N2}$ (Figure 5c). The responsivity is further improved by the combination of positive gate biasing and light illumination, which cause both the capacitively- and photo-induced electrons to accumulate (Figure 5d). Consequently, a huge $\Delta R/R_{N2}$ enhancement is accomplished. The oxygen sensing performance in this study appears to be particularly noteworthy among reported sensors, which are summarized in supplementary **Table S1**.

The stability and durability of the ReS$_2$ FET-based oxygen sensor properties are also vital for practical applications. To examine these factors, measurements were repeated in light illumination and positive gate voltage operations. The device was exposed to oxygen gas every week for 35 days in dry or humid conditions. As shown in **Figure 6**a and 6b, the responsivity at an oxygen concentration of 10,000 ppm was retained after 35 days of measurement and it was largely unchanged even when measured in humid conditions. That is, our sensor showed long-term stability, and stable operation even under humid conditions. We suppose that this high stability is brought about by the high crystallinity of the ReS$_2$ nanosheet. These results suggest that our sensor could be employed in real-time applications.

In conclusion, we have developed high-performance ReS$_2$ FET-based oxygen sensors by combining light illumination and a gate biasing operation. The underlying mechanism was explained by the electron transfer from ReS$_2$ into oxygen molecules to stimulate changes in the transistor properties. Light illumination and gate biasing improved the oxygen sensing properties by increasing the photogenerated carrier and doping level of the ReS$_2$ FET. In consequence, a practical sensitivity of 0.01% ppm$^{-1}$ was achieved, which outperform results from the previous reports. Furthermore, long-term stability, and stable operation even under humid conditions enabled the sensor device to be used in real-time applications. These results imply that a light-assisted and gate-bias-tunable oxygen sensor based on an ReS$_2$ FET has the potential for allowing further sensor development towards versatile tunable oxygen sensors.

**Experimental Section**

*Preparation and characterization of ReS$_2$ nanosheets*: ReS$_2$ nanosheets were prepared by a gold-mediated exfoliation method,[35] which is effective for preparing large-scale ReS$_2$ nanosheets with a uniform thickness. With this method, a bulk ReS$_2$ crystal (HQ Graphene supplier) was mechanically exfoliated on a thermal tape (Nitto Denko, model NO319Y-4LSC). A 100 nm-thick gold film was then evaporated directly onto an as-exfoliated ReS$_2$/thermal tape. Another thermal tape was used to exfoliate Au/ReS$_2$ layers from the initial ReS$_2$/Au coated thermal tape, and then the exfoliated Au/ReS$_2$ layers were pasted onto SiO$_2$/Si substrates. Next, the thermal tape was detached at a temperature of 100°C. Finally, the substrate was immersed in Au etchant (AURUM) for four minutes and then rinsed in deionized water for five minutes to remove the residual gold layer from the ReS$_2$ surface. The crystallinity and film thickness of thus prepared ReS$_2$ nanosheets were characterized by Raman microscopy (Nanophoton, model: Ramanplus) and AFM. To avoid any possible effect of heating on the ReS$_2$ nanosheets, the Raman laser power on the sample was fixed at 1 mW. The fabrication process and an optical image are shown in **Figure S4**.

*Fabrication of ReS$_2$ FET-based oxygen sensor*: To fabricate the FET-based sensor, a SiO$_2$/Si substrate with ReS$_2$ nanosheets was coated with a PMMA layer and patterned by using electron-beam lithography (Elionics, ELS-7000), followed by Cr/Au (3 nm/80 nm) metal deposition using e-beam evaporation and a lift-off process. The ReS$_2$ channel length and width were 0.65 µm and 2 µm, respectively. Here, the highly-doped Si substrate and the 285 nm-thick SiO$_2$ layer worked as a gate electrode and a gate dielectric layer, respectively. The sample device was then cut to form a 2.5 mm × 2.5 mm chip by using a manual scribing machine and was mounted on an 8-pin standard chip carrier using silver paste. A manual wedge-wedge bonder was utilized to realize a wire-bond connection between the chip carrier and the sensor device.

*Sensing measurement of ReS$_2$ FET-based oxygen sensor*: The home-made gas measurement setup was equipped with a mixing connector linked to mass flow controllers that made it possible to adjust the oxygen concentration by changing the $O_2/N_2$ gas mixing ratio. A constant flow rate of 100 mL/min was used for dry air and oxygen gas. The gas chamber was flushed with nitrogen for a few hours to recover the device to the initial state after each measurement. The setup was also equipped with a relative humidity (RH) controller. The RH levels were monitored with a commercial humidity sensor. All the current-voltage (I-V) measurements were carried out at room temperature using a source-measurement unit (Keysight B2912A), which was connected to a data acquisition system using EasyEXPERT software. A Xenon lamp (Asahi spectra Co. Ltd., MAX 303) was employed for light irradiation. The sensor was illuminated with red light at a wavelength of 650 nm (**Figure S5**) through the acrylic resin window of the gas sensor chamber. The light intensity was monitored with a power meter (Ophir Optics, PD300).

**Supporting Information**

Supporting Information is available from the Wiley Online Library or from the author.

**Acknowledgements**

This research was supported by the World Premier International Center (WPI) for Materials Nanoarchitectonics (MANA) of the National Institute for Materials Science (NIMS), Tsukuba, Japan. A part of this study was supported by a Grant-in-Aid for Scientific Research (JSPS KAKENHI Grant No./Project/Area No.17F17360), and the NIMS Nanofabrication Platform and the NIMS Molecule & Material Synthesis Platform in the Nanotechnology Platform Project sponsored by the Ministry of Education, Culture, Sports, Science, and Technology (MEXT), Japan.

Received: ((will be filled in by the editorial staff))
Revised: ((will be filled in by the editorial staff))
Published online: ((will be filled in by the editorial staff))

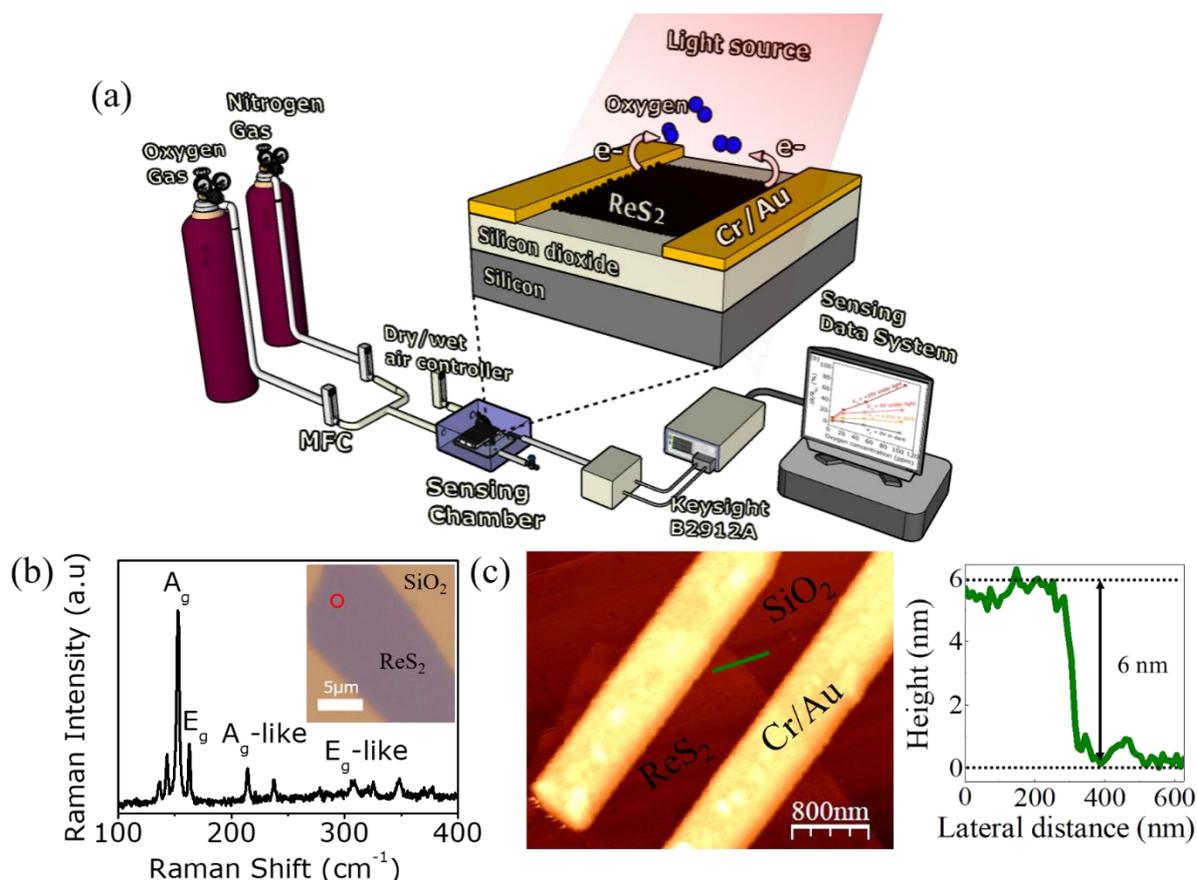

**Figure 1.** (a) Schematic diagram of the home-made gas sensing measurement setup and the ReS$_2$ FET-based oxygen sensor. (b) Raman spectrum of the ReS$_2$ nanosheet obtained at the location indicated by the red circle seen in the inset optical image. (b) AFM image and height profile of the 8-layer ReS$_2$ nanosheet.

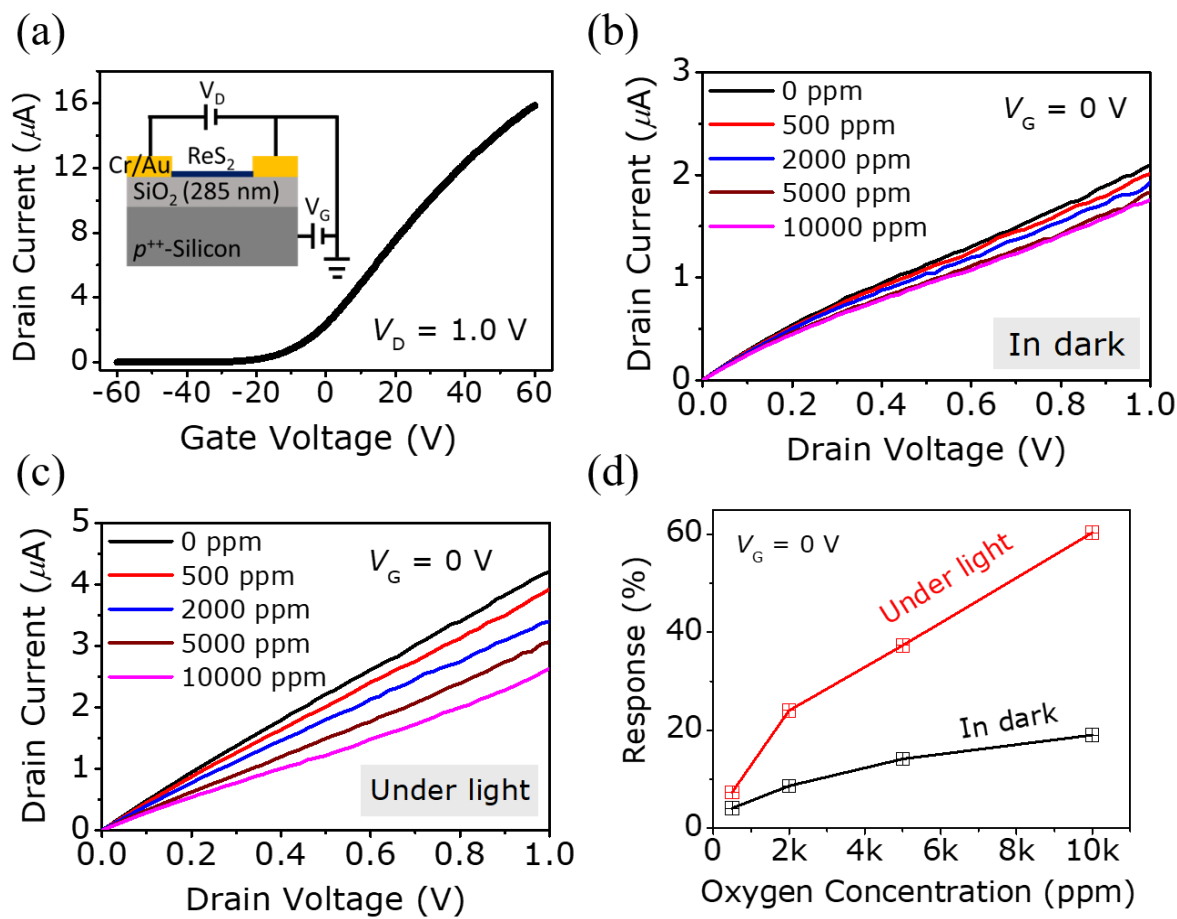

**Figure 2.** (a) Transfer curve of the 8-layer ReS$_2$ FET, and its output characteristics (b) in the dark and (c) with light illumination. The O$_2$ gas concentration ranged from 0 to 10,000 ppm. No gate voltage was applied. (d) Responsivity as a function of O$_2$ gas concentration in the dark (black line) and with illumination (red line).

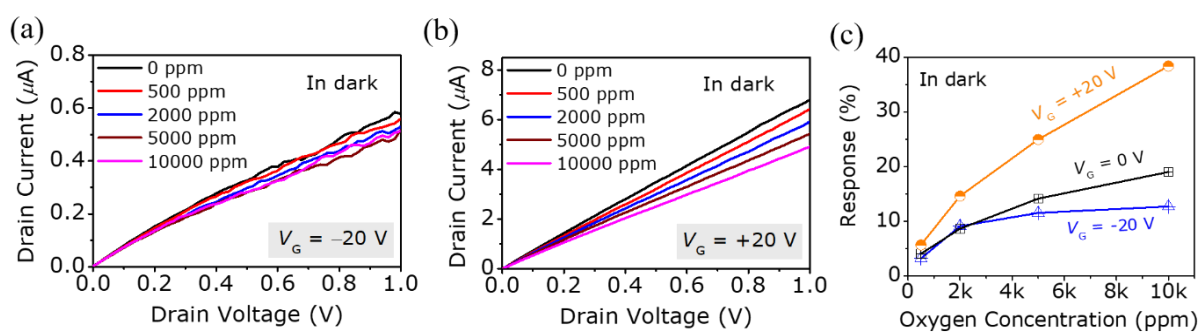

**Figure 3.** Output characteristics of the 8-layer ReS$_2$ FET under (a) $V_G = -20$ V, (b) $V_G = +20$ V and O$_2$ gas exposure at concentrations of 0 to 10,000 ppm measured in dark conditions. (c) Responsivity as a function of O$_2$ gas concentration curves of the sensor with and without applying a gate voltage in dark conditions.

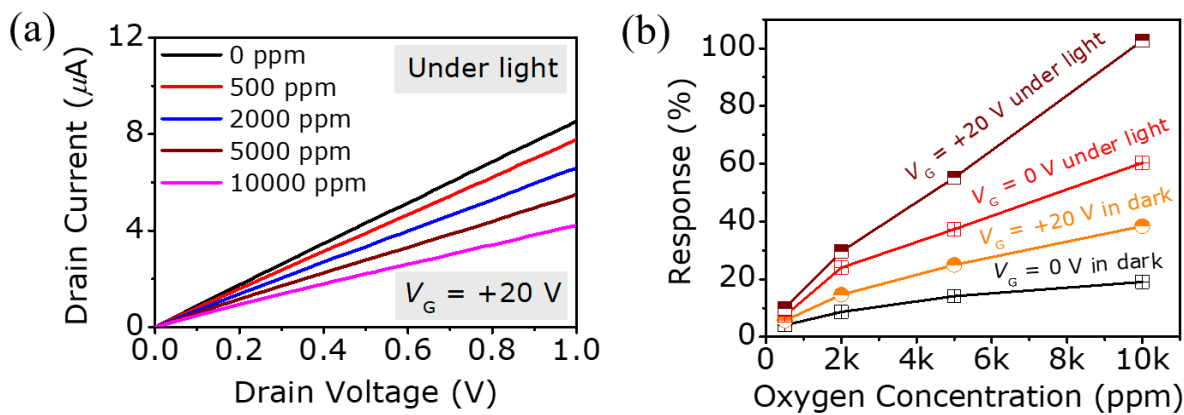

**Figure 4.** (a) Output characteristic of the 8-layer ReS$_2$ FET where $V_G$ = +20 V and in an O$_2$ gas atmosphere at various concentrations ranging from 0 to 10,000 ppm measured under light illumination. (b) Responsivity as a function of O$_2$ gas concentration when operated under various conditions: with/without gate voltage and with/without light illumination.

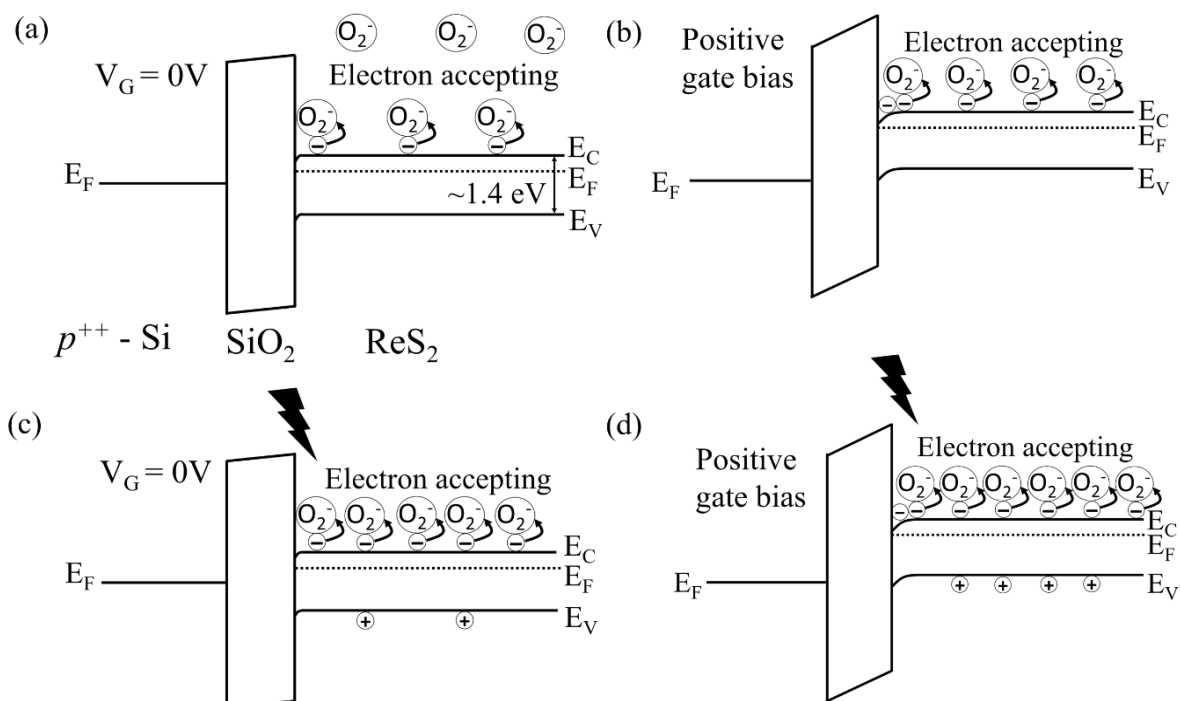

**Figure 5.** Band diagrams of the ReS$_2$ FET, illustrating the interaction of electrons in the ReS$_2$ with oxygen gas (a) under dark conditions without a gate bias, (b) under dark conditions with a positive gate bias, (c) under light illumination without a gate bias, and (d) under light illumination with a positive gate bias.

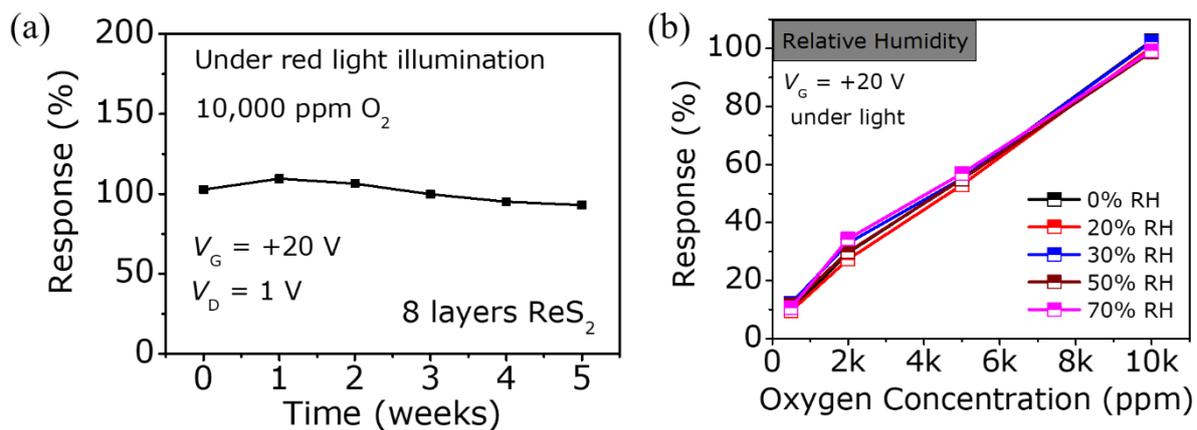

**Figure 6.** (a) Long-term stability test at 10,000 ppm O$_2$ gas concentration. (b) Influence of relative humidity on device responsivity. The relative humidity was varied from 0% to 70% by controlling the mixture ratio of dry and wet air.

Rhenium disulfide (ReS$_2$) field-effect transistor (FET) based on a light-assisted and gate-voltage operation as a potential high-performance oxygen gas sensor. Demonstration of a practical sensitivity via the combined roles of light illumination and a gate biasing operation. Key elements are a photogenerated carrier and the doping level of the ReS$_2$ FET, which increased for interaction with oxygen molecules. Long-term stability and stable operation under humid condition. Here, a crystalline ReS$_2$ nanosheet was effectively utilized. The underlying mechanism is explained in terms of the electron doping from the ReS$_2$ to the oxygen molecules, which stimulated changes in the transistor characteristics. Controllability of the light illumination and gate voltage and, therefore, ReS$_2$ FETs can play an important role in a high-performance oxygen sensor.

**Keyword** ReS$_2$, field-effect transistors, oxygen sensors, light-assisted, gate tunability

Amir Zulkefli, Bablu Mukherjee, Ryoma Hayakawa, Takuya Iwasaki, Shu Nakaharai\*, Yutaka Wakayama\*

**Light-Assisted and Gate-Tunable Oxygen Gas Sensor based on Rhenium Disulfide (ReS$_2$) Field-Effect Transistors**

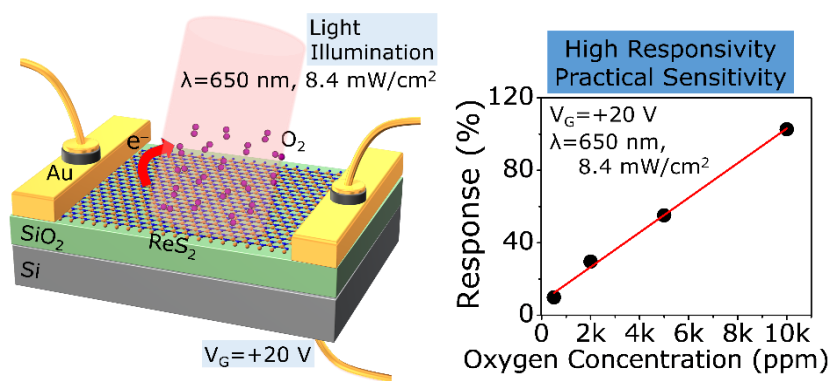

Supporting Information

**Light-Assisted and Gate-Tunable Oxygen Gas Sensor based on Rhenium Disulfide (ReS$_2$) Field-Effect Transistors**

*Amir Zulkefli, Bablu Mukherjee, Ryoma Hayakawa, Takuya Iwasaki, Shu Nakaharai\*, Yutaka Wakayama\**

Influence of ReS$_2$ thickness on oxygen sensing response

To clarify the role of ReS$_2$ thickness, 6-, 8- and 11-layer-thick ReS$_2$ FETs were prepared. The response of each sensor was examined under oxygen concentrations varying from 0 to 10,000 ppm as shown in Figure S1. All the gas sensing measurements were conducted out under light illumination and without a gate voltage. As the thickness was increased to 11 layers, the oxygen response of the device provided comparable data to that of an 8-layer ReS$_2$ FET, which suggested that 8-layer-thick ReS$_2$ is sufficient for use in a high-performance oxygen sensor.

This could be due to any of many factors such as the trade-off dependence between the surface-to-volume ratio[1] and the optical absorbance[2] on the light-assisted gas sensing properties. A high surface-to-volume ratio favors more gas adsorption on the channel surface and accelerates the charge accumulation to interact with gas molecules.[3] Meanwhile, the optical phenomena of few-layer ReS$_2$ (8 layers) is sufficient to obtain reasonable optical absorption,[4] which we assume to provide an optimal optoelectrical response to the sensors compared with a monolayer in order to efficiently collect photogenerated carriers at the top electrodes.

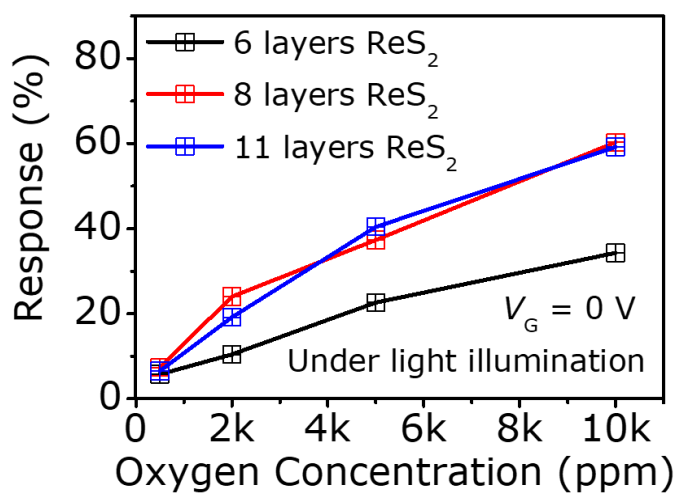

**Figure S1.** Responsivity as a function of oxygen gas concentration (electron accepting behavior) for 6-, 8-, and 11-layer ReS$_2$ FETs.

Influence of light intensity on oxygen sensing response

To clarify the influence of light intensity on the oxygen sensing response, an 8-layer ReS$_2$ FET was exposed to oxygen gas concentrations ranging from 0 to 10,000 ppm under illumination with four different light intensities: 1.6, 5.6, 8.4 and 12.0 mW/cm$^2$, as shown in Figure S2. It is apparent that the optimal responsivity toward oxygen gas was achieved at 8.4 mW/cm$^2$ by comparison with that at 1.6, 5.6, and 12 mW/cm$^2$.

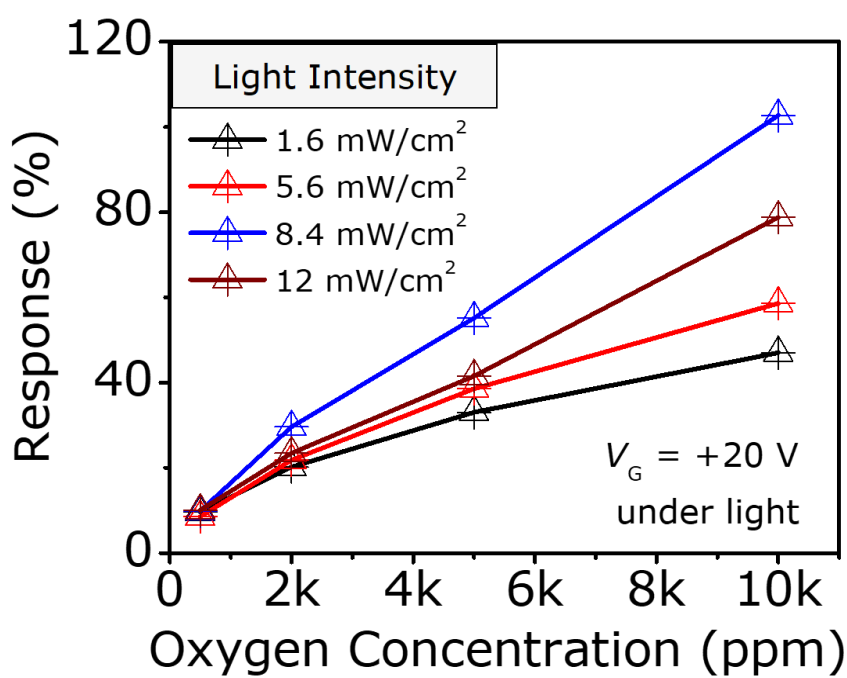

**Figure S2.** Influence of light intensity variation on 8-layer ReS$_2$ FET sensor device responsivity.

The oxygen gas sensitivity ($S_{O2}$) was obtained from the slope of the line between the responsivity and oxygen concentration by using the following equation:

$$S_{O2} = \left[\frac{\partial\,(\Delta R/R_{N2})}{\partial\,C_{O2}}\right] \quad (2)$$

where $C_{O2}$ represents oxygen concentration.

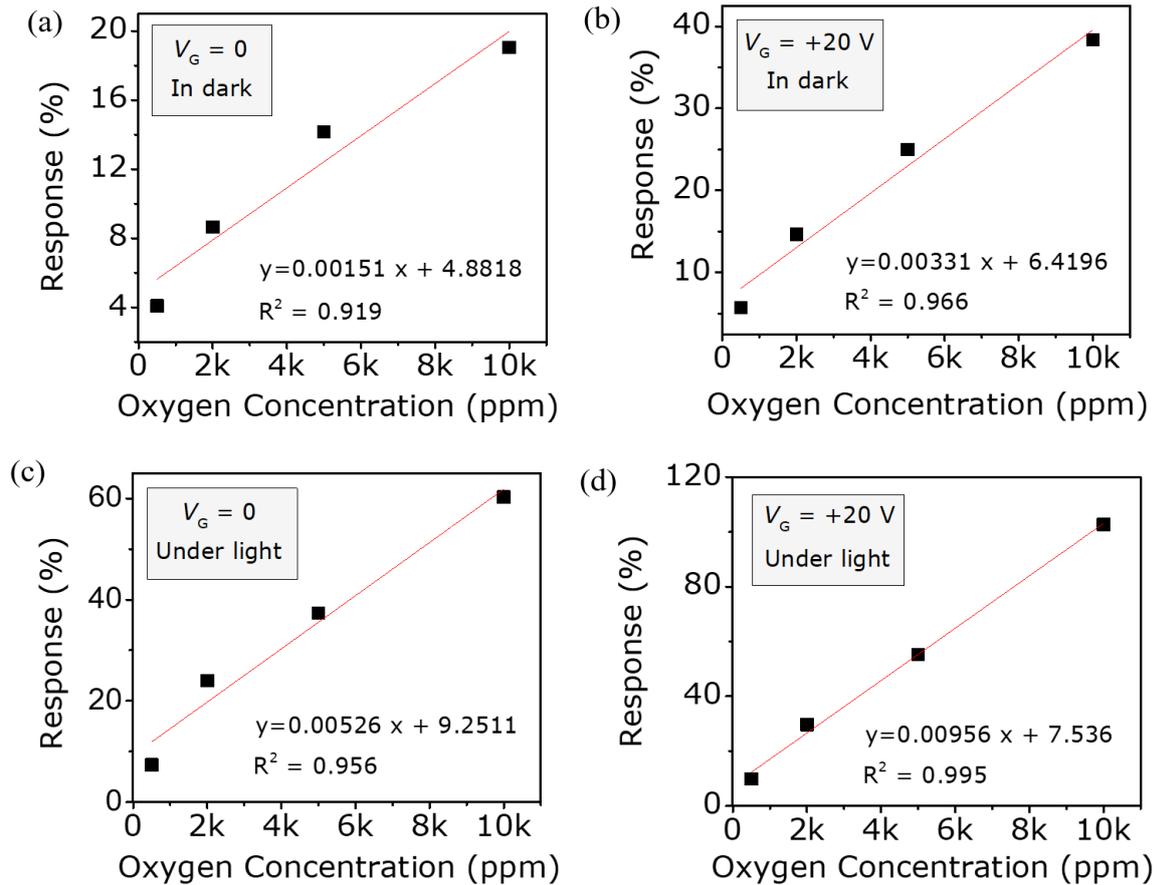

**Figure S3.** Linear fit curves of the responsivity value as a function of oxygen concentration under different operating conditions: (a) in the dark without a gate bias, (b) in the dark with a +20 V gate bias, (c) under light without a gate bias, and (d) under light with a +20 V gate bias. The sensor responsivity exhibited linear behaviors. These behaviors show that the sensor device can be sensitively and reliably operated over a wide concentration range.

Table S1 summarizes the oxygen sensing properties of different oxygen gas sensors described in previous reports. Our oxygen gas sensor, which we operated under a combination of light illumination and a +20 V gate voltage, is noteworthy in terms of responsivity and sensitivity as compared with the earlier reports.

**Table S1.** Oxygen gas sensing properties of different oxygen gas sensors in the previous studies.

| Sensing material | Device type | Oxygen concentration | Response | Sensitivity | Operating temperature | Ref. |
|---|---|---|---|---|---|---|
| SrTiO$_3$ | Chemiresistor | 20% | 6.35 ($R_{O2}/R_{N2}$) | --- | 40°C | [5] 2004 |
| Pt-In$_2$O$_3$ | Chemiresistor | 20% | 95% ($\Delta R/R_{N2}$) | --- | RT | [6] 2005 |
| Pt-In$_2$O$_3$ | Chemiresistor | 20% | 63.3% ($\Delta R/R_{N2}$) | --- | 200°C | [7] 2007 |
| Graphene/TiO$_2$ | Chemiresistor | 134 ppm | 7% ($\Delta R/R_{N2}$) | --- | RT | [8] 2013 |
| ZnO | Chemiresistor | 97% | 419% ($\Delta R/R_{N2}$) | 1.83%/%O$_2$ | 50°C / UV assisted | [9] 2014 |
| MWCNTs/PVAc/TiO$_2$ | Chemiresistor | 5% | 30 ($R_{O2}/R_{N2}$) | --- | 400°C | [10] 2014 |
| MoS$_2$ | FET | 9.9x10$^{-5}$ millibars | 29.2% ($\Delta I/I_{N2}$) | 26.7x10$^4$%/mBar | RT / +40V gate-assisted | [11] 2015 |
| MoS$_2$ | Chemiresistor | 100% | 63.73% ($\Delta R/R_{N2}$) | 0.545%/%O$_2$ | 300°C | [12] 2016 |
| TiO$_2$ | Chemiresistor | 1% | 1.15 ($R_{O2}/R_{N2}$) | --- | RT | [13] 2016 |
| TiS$_2$ | Chemiresistor | 1% | 34.8% ($\Delta R/R_{N2}$) | 1.75%/1.0% | RT | [14] 2020 |
| ReS$_2$ | FET | 10,000 ppm | 102.68% ($\Delta R/R_{N2}$) | 0.01%/ppm | RT / light-assisted and gate-bias assisted | This work |

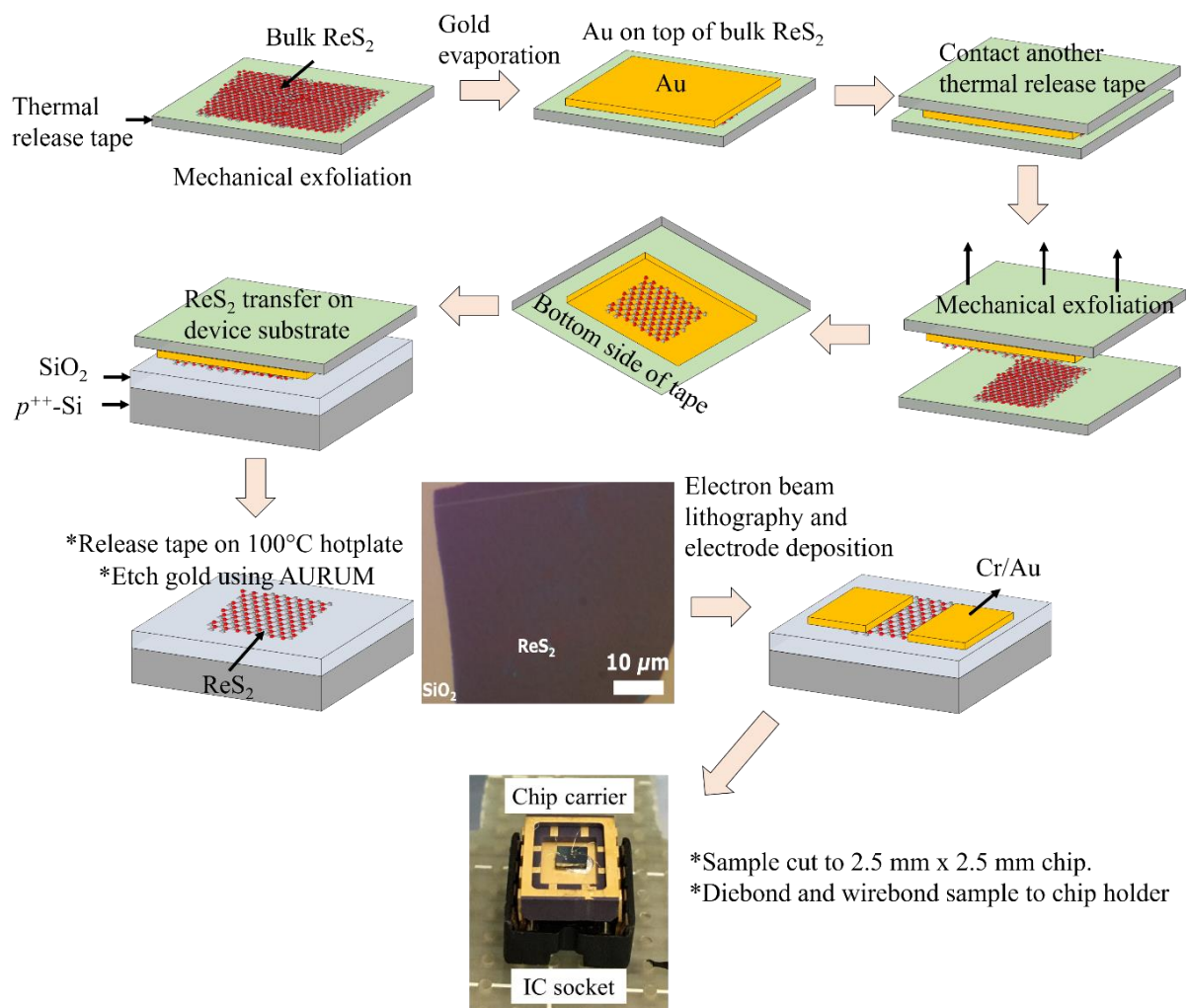

**Figure S4.** Preparation of the ReS$_2$ nanosheets by using gold-mediated exfoliation and a photograph of the ReS$_2$ FET-based oxygen sensor mounted on a chip carrier.

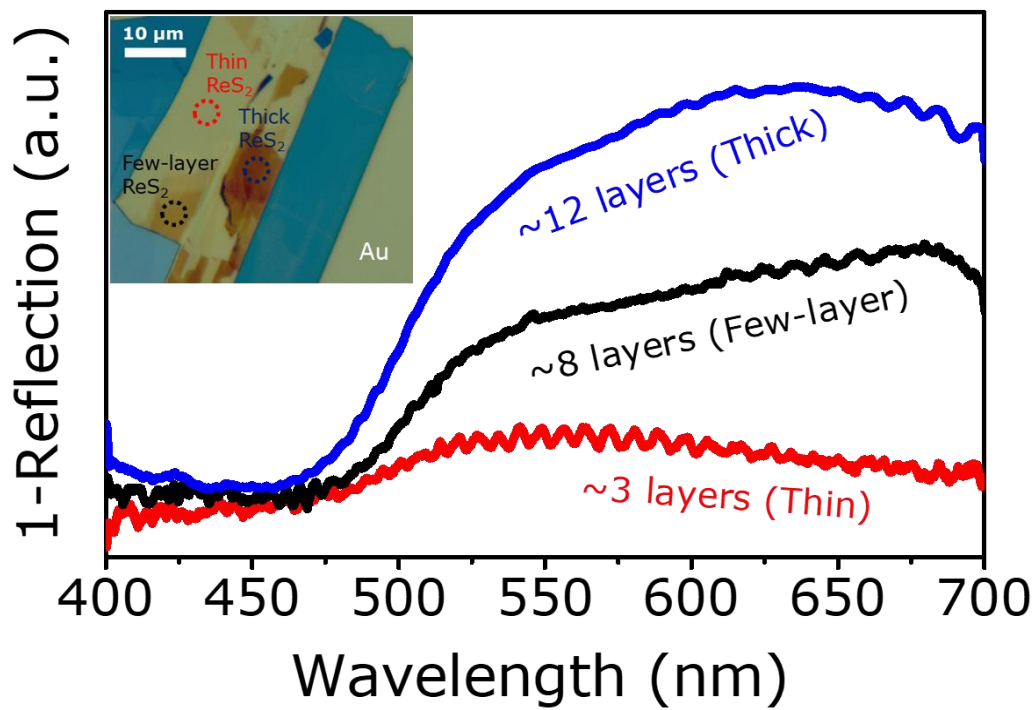

**Figure S5.** Absorption spectra (1-Reflection) of ReS$_2$ nanosheets. Inset is an optical image showing thin-layer, a few-layer and thick-layer ReS$_2$ on top of a Au film coated substrate. Light illumination of 650 nm wavelength was utilized for oxygen sensing measurement of 8-layer ReS$_2$ FET, which observed to provide reasonable absorption.